\documentclass[pra,twocolumn,longbibliography]{revtex4-2}
\usepackage[utf8]{inputenc}
\usepackage{units}
\usepackage{amsmath}
\usepackage{amssymb}
\usepackage{mathtools}
\usepackage{graphicx}
\usepackage{float}
\usepackage[dvipsnames]{xcolor}
\begin{document}
\title{%
Perspective on electromagnetically induced transparency vs Autler-Townes splitting}
\author{Barry C.\ Sanders}
\affiliation{Institute for Quantum Science and Technology, University of Calgary, Alberta T2N~1N4, Canada}
\email{sandersb@ucalgary.ca}
\date{\today}
\begin{abstract}
Electromagnetically induced transparency and Autler-Townes splitting are two distinct yet related effects.
These phenomena are relevant to  quantum technologies,
including quantum memory, quantum switching
and quantum transduction.
Here we discuss the similarities and differences between these phenomena along historical and conceptual lines
and elaborate on their realisations on various physical platforms
including atomic gases,
superconducting circuits and optomechanics.
In particular, we clarify two  approaches to assessing which phenomenon is observed based on a
black-box approach of modelling output given particular input vs analysing the underpinning physics.
Furthermore, we highlight the ability to effect a continuous transition between the two seemingly disparate phenomena.
\end{abstract}
\maketitle
\section{Introduction}
This article is based on a presentation I gave at the Jonathan P.\ Dowling Memorial Conference 9--11 January 2023 at the Quantum Terminal in Central Station located in Sydney, Australia.
This conference provided an opportunity to mourn Dowling's passing~\cite{FW20}
and to reflect on his significant contributions to physics.
This article reflects on collaborative work with Dowling and his at-the-time postdoctoral research associate P\"{e}tr Anisimov~\cite{ADS11}.
Our collaboration commenced at the 4th
International Conference titled ``Frontiers of Nonlinear Physics'',
which took place entirely on a
boat
travelling on the Volga river from Nizhny Novgorod to St.~Petersburg during the period
13--20 July 2010
(followed by months of fine-tuning the analysis and the message before completing the manuscript and submitting to the journal).
Many experts on electromagnetically induced transparency (EIT) were on the boat with us,
which afforded the opportunity to explore and debate the fundamentals of EIT
vs its counterpart Autler-Townes splitting (ATS).

The context for EIT is controlling coherent processes in atoms and molecules,
including coherent population trapping~\cite{Ari96},
lasing without inversion~\cite{Koc92} and EIT~\cite{BIH91,Mar98,FIM05}.
Although originally performed with alkali gases~\cite{FIM05} in an electronic~$\wedge$ configuration
(often written~$\Lambda$ and called Lambda),
EIT has been reported in other systems,
including 
quantum dots~\cite{XSB+08},
metamaterials~\cite{PFZP08,ZGW+08},
nanoplasmonics~\cite{LLW+09},
superconducting circuits~\cite{KDS+10},
and optomechanics~\cite{SAC+11}.
Interestingly,
EIT is achieved for classical coupled oscillators, 
in particular for inductively or capacitively coupled electrical resonator circuits sharing a common environment~\cite{LR51,GMN02}.
EIT is germane for
exploiting constructive interference of the nonlinear susceptibility whilst achieving destructive interference of the linear susceptibility~\cite{HMS91,Har97},
all-optical fast switching~\cite{HY98},
optically controlled slowing of light~\cite{HHD+99,FL00}
and optical quantum memory~\cite{PFM+01,LST09,SAA+10}.

The concept of ATS was introduced in a 1955 manuscript by Autler and Townes on
``Stark effect in rapidly varying fields''~\cite{AT55}.
The Autler-Townes effect is often called the dynamic, or ac, Stark effect, distinct from the original dc Stark effect, which
refers to splitting of lines in the presence of an electric field,
analogous to Zeeman splitting for the magnetic field~\cite{Kox13}.
Despite ``Stark'' being mentioned in both dc and ac cases,
the ATS ac case is novel and different from the original Stark shift.
Physically, ATS is about
splitting a spectral line
by a field that is at or nearly on resonance with the transition frequency.
In the words of Cohen-Tannoudji~\cite{Coh96}
\begin{quote}
a \dots\ transition \dots\ can split into two components [Autler-Townes doublet] when one of the two levels involved in the transition is coupled to a third one by a strong resonant \dots\ field
\end{quote}
with this splitting arising due to multiphoton interference in absorption of the driving fields.
Prima facie, ATS is quite different physically from EIT.
\section{Discerning electromagnetically induced transparency from Autler-Townes splitting}
The impetus for my project with Anisimov and Dowling on discerning EIT from ATS
arose from the claim of EIT in superconducting circuits under noisy conditions~\cite{AAZ+10}.
This beautiful experimental result demonstrated fully controllable near-perfect control of reflection and transmission of propagating microwaves through artificial atoms.
Particularly interesting for us was their use of a three-level cascade system, one which I denote with the symbol~$\Xi$ to resemble a ladder,
rather than the widely used~$\wedge$ system used for EIT.

Unfortunately, this superconducting-system experimental signature of EIT was quite noisy~\cite{AAZ+10}.
What they had shown was, without a doubt, control of reflection and transmission, hence EIT in a literal sense;
however, whether they had observed EIT in the conventional sense of observing transparency that is induced coherently even if the pump field is arbitrarily weak was in doubt.
Note that not all experiments at that time were so noisy;
the experimental result of optomechanically induced transparency,
which achieves induced transparency via radiation-pressure coupling between optical and mechanical modes,
showed EIT meticulously in the conventional sense~\cite{WRD+10}.

Many world experts were on the Volga cruise,
but I had the luck of talking to Dowling and Anisimov about the puzzle of whether EIT can be discerned under noisy conditions and with a not-so-weak driving field.
Fortuitously,
Anisimov had co-authored a vital paper with Olga Kocharovskaya
on decaying-dressed-state analysis of a coherently driven three-level $\wedge$ system,
which explains that the difference between EIT and ATS~\cite{AK08}
\begin{quote}
originates from the difference of two Lorentzians centered at the same position, rather than the summation of two Lorentzians shifted by twice the Rabi frequency.
\end{quote}
They relate the heart of this stark difference between EIT and ATS to the well known phenomenon of Fano interference~\cite{Fan61} between two excitation pathways~\cite{FIM05};
Fano interference was recognised as key to the related lasing-without-inversion phenomenon somewhat earlier as well~\cite{Har89}.
In subsequent, independent work on EIT vs ATS,
Abi-Salloum explains that~\cite{Abi10}
\begin{quote}
 the [EIT] dip is a result of a destructive interference between the two resonances [and] an ``imprint'' of one resonance into the other.
\end{quote}
Abi-Salloum's view is compatible with Fano interference being at the heart of the dip.

The role or absence of observable
Fano interference can be understood as follows.
The difference between two Lorentzians centred at the same point is a form of Fano interference arising due to a shared reservoir.
In contrast, the sum of two mutually displaced Lorentzians does not exhibit Fano interference:
the splitting of the line is a pumping induced formation of a doublet structure in the absorption profile.

EIT and ATS are hard to distinguish for strong driving fields or for noisy systems but easy to distinguish for the low-noise weak-driving limit.
Anisimov and Dowling and I then realised that the task of identifying whether EIT holds or not comes down to hypothesis testing:
whether Fano interference is observed or not,
and indeed allowing a continuous transition from EIT to ATS.
Mathematically, the Kullback-Leibler divergence,
or relative entropy, is useful as a quantifier by how surprising the alternative would be.
However, fitting parameters are used in EIT and ATS models so we have to penalise fitting parameters.
This penalisation is achieved by employing (a modified) Akaike's information criterion, which~\cite{BA02}
\begin{quote}
identifies the most informative model based on Kullback-Leibler divergence (relative entropy), which is the average logarithmic difference between two distributions with respect to the first distribution.
\end{quote}
Thus, we pose mathematically the task of deciding whether the claim that EIT is observed by quantifying the surprise if it isn't,
and a lower level of surprise, as in the noisy or strong-driving cases, indicates that the experiment is less convincing.
\section{Classical or quantum phenomena}
Now we address how `quantum' EIT and ATS are.
Are these two phenomena classical in nature?
The first hint of EIT arguably goes back to the work of Lamb and Retherford in 1951~\cite{LR51},
which improves on their landmark ``Lamb shift'' observation~\cite{LR47}.
They study the fine structure of Hydrogen,
specifically the shift of the~2$^2$S$_{\nicefrac12}$ level.
The resonances that they show, such as in Fig.~45 of their paper, are explained in terms of their Fig.~46,
which depicts two coupled harmonic oscillators manifested as inductor-capacitor circuits.
Only one of the two circuits forming their analogy is driven by an alternating-current source;
that same circuits also has a resistor.
The other, undriven circuit does not have a resistor and is inductively coupled to the first.
This pair of coupled circuits thus shares a single reservoir, due to the single resistor that jointly damps both.
The point I am making is that their understanding of an EIT type of phenomenon is strictly classical and strikingly clear and compelling.

In independent work half a century later, in 2002,
a classical analogue of EIT is explained in terms of
coupled oscillators~\cite{GMN02}.
This paper advocates the classical analogue as a pedagogical approach to teaching EIT and demonstrates with a pair of coupled resonant circuits akin to the intuition of Lamb and Retherford.
One can draw the conclusion that EIT is classical in a phenomenological sense:
what is seen can be explained classically even if the underlying physics is known to be quantum in the sense of involving discrete energy levels of a three-level system.
One way to understand this point is to regard testing EIT vs ATS as being classical as a black box
(maps input to output with the intermediate procedure unknown/unrevealed), which is understandable classical,
vs white box
(reveals the inner workings)
wherein quantum mechanics is needed to explain that \emph{particular} procedure involving, say, three-level atoms.

Thinking in a black-box way enables us to define analogues to EIT that exhibit absorption and dispersion profiles akin to those of EIT with the same kind of controllability.
I mentioned EIT in superconducting circuits~\cite{AAZ+10} already and the problem of discerning EIT from ATS.
Although that result, at the time,
somewhat blurs whether EIT or ATS is present,
not every experiment suffers this drawback.
A contemporaneous realisation of optomechanically induced transparency (abbreviated as OMIT and is transparency induced through coupling of an optical mode with a mechanical mode through radiation pressure),
which employs an optical cavity with one cavity mirror attached to a mechanical oscillator~\cite{WRD+10}.
The system is operated at a sufficiently low temperature to enable resolution between different vibrational modes of the oscillator.
In this system,
the weak optical probe-field response matches what is expected of a probe response for EIT.
\section{Implications and applications}
The idea of discerning EIT from ATS has caught on because determining which phenomenon pertains reveals whether the underlying physics is a Fano interference or line splitting due to a strong driving field.
The first experimental study of the EIT-to-ATS transition employed our modified Akaike's information criterion
to their controlled system of cold c\ae sium atoms~\cite{GVS+13}.
One benefit of applying this test to their system was the clear revelation of strong sensitivity to the properties of the medium,
thus ``potentially providing a practical characterizing tool''~\cite{GVS+13}.

An interesting theoretical study suggests that an EIT-to-ATS transition is possible for a~$\vee$ configuration in hot molecules, thus extending from the previously mentioned~$\wedge$ and~$\Xi$ configurations~\cite{ZTH13}.
Additionally, they claim that the EIT-to-ATS transition is not allowed for cold molecules.
They show that their results compare favourably with an experimental demonstration of EIT in inhomogenously broadened Na$_2$ molecules~\cite{LKA+11}.

The types of systems showing EIT-to-ATS transitios is quite varied and rich.
Controlled the transition from EIT
and ATS has been demonstrated in a realisation of
coupled whispering-gallery-mode resonators~\cite{POC+14}.
They control the transition by varying resonator separation thereby increasing the coupling strength,
and the stage of the transition from EIT to ATS is quantified by our modified Akaike's Information Criterion~\cite{ADS11}.
Another case of the EIT-to-ATS transition has been demonstrated for coupled mechanical oscillators system~\cite{LYW+16}.
Suggestions have been made to see this transition in plasmonic waveguides~\cite{HWG+15}.

The EIT-to-ATS transition from EIT to ATS has been shown for~$\Xi$ configurations in cold atoms.
One demonstration concerned the c\ae sium atoms involving the 35S$_{\nicefrac12}$ Rydberg level~\cite{HJX+18}.
Another demonstration of the EIT-to-ATS transition was achieved for the~$\Xi$ configuration of~$^{87}$Rb,
achieved by the transition
5S$_{\nicefrac12}\to$5P$_{\nicefrac32}\to$5D$_{\nicefrac52}$~\cite{NNCR19}).

One important application of EIT is optical quantum memory,
but explorations of the EIT-to-ATS transition opened a new way of thinking about quantum memory.
Coherent storage and control of broadband photons is possible without EIT by instead dynamically controlling ATS~\cite{SHR+18,RSH+19}.
They point out that EIT-based quantum memory adiabatically eliminates absorption in contrast to ATS being based on absorption~\cite{RSH+19}:
\begin{quote}
We find that their storage characteristics manifest opposite limits of the light-matter interaction due to their inherent adiabatic versus nonadiabatic nature.
\end{quote}
They advocate EIT storing narrow-band fields and ATS protocol is intrinsically suited for storing broad-band fields.

Thinking about EIT vs ATS in the context of~$\wedge$, $\Xi$ and~$\vee$ systems
helped to reintroduce~$\Delta$ systems,
in which each pair of the three levels is coherently coupled.
The underlying physics of coherently coupling all transitions,
say by adding a $\mu$wave field to couple two transitions with the other optical transitions,
had been studied for quite some time theoretically~\cite{DK82,KMR91,KK00}
and experimentally~\cite{ZWH+97,WMWL05}.
In particular,
the paper by Kosachiov and Korsunsky in the year 2000 is quite relevant for our discussion but distinct:
they consider $\mu$wave conversion between the two lower levels in two regimes, EIT for a weak field and what they call ``Autler-Townes-type EIT'' for the strong field~\cite{KK00}.

Moreover,
the~$\Delta$ configuration could yield EIT plus amplification~\cite{JBBS10},
and high-contrast optical switching has been shown for the~$\Delta$ system~\cite{GKJ+17}.
Separating EIT and ATS types of effects are not fully explored yet but would essentially be about testing for observable Fano interferences in the system.

Another system ripe for studying EIT vs ATS arises for double-EIT systems~\cite{RVO+04}, and their analogues such as double-OMIT~\cite{LMSB20}.
Further extensions exist such as double-double EIT for which each of the signal and probe fields have two transparency windows~\cite{AS14,AS15}.
Exploring how much Fano interference is needed for applications of these extensions to EIT and whether ATS alternatives suffice would be interesting.
\section{Conclusions}
Finally,
I now discuss the issue of whether EIT and ATS are linear or nonlinear optical phenomena.
Discerning EIT from ATS is based on deciding whether the dispersion or absorption profile,
based on linear-optical intuition,
shows Fano interference or not.
However, EIT is fundamentally a nonlinear~$\chi^{(3)}$ optical effect~\cite{PSZ95} with two fields involved:
a pump field and a probe field.
The linear-optical simplification is achieved by fixing the strength of the pump field and incorporating this fixed value into the calculation of an effective first-order, or linear,
susceptibility,
i.e., a linear-response limit,
which I call ``pseudo-linear optics''
to make clear that the dynamics are not truly of the linear-optical type.
However, fixing the pump field to zero strength would of course remove EIT so the phenomenon of EIT is fundamentally nonlinear but conveniently cast into a linear-optical framework, and the same for ATS.
Furthermore, nonlinear optical effects arise in the EIT setting under appropriate conditions~\cite{KK86}.

Let us understand better this  effective pseudo-linear-optical description of EIT,
which means in effect that we can discuss meaningfully (linear) absorption and dispersion despite having two driving fields.
Essentially, this pseudo-linear treatment arises by dressing the states,
which builds into their descriptions the interference that occurs from these two driving fields~\cite{Coh96}.
Such dressed states are obtained by
field-modified coherent population trapping~\cite{GWS78}.
Another way of viewing this coherent population trapping is as the formation of a dark state in a~$\wedge$ electronic configuration.
This dark state is an anti-symmetric coherent superposition of the two sublevel states with constraints on the weights of these two states.
Specifically,
these weights scale inversely with their respective Rabi frequencies,
thus ensuring destructively interfering transitions to the upper state leading to no absorption.

{\bf Acknowledgments:}
This work has been supported by Canada's Natural Sciences and Engineering Research Council (NSERC).
I acknowledge enjoyable and valuable discussions on EIT and ATS with Tony Abi-Salloum, P\"{e}tr Anisimov, Andal Narayanan and Olga Kocharovskaya.
Furthermore, I appreciate Jonathan Dowling's role in our joint work, his insights, his wit, his gregarious outreach activities that helped to get our work known to the wide quantum optics community and, above all, his friendship.

{\bf Data Availability:}
Data sharing is not applicable to this article as no new data were created or analyzed in this study.

{\bf Conflict of interest:}
The author has no conflicts to disclose.
\bibliography{eitats}
\end{document}